\documentstyle{article}

 \normalmarginpar
\twocolumn
 \makeatletter
 \gdef\@proofbox{\relax}
 \long\def\proofbox#1{\gdef\@proofbox{#1}}
 \proofbox{\small{\sl PhysComp96}\\\ifx\UNDEF\@fullpaper
     Extended abstract\else Full paper\fi\\Draft, \@date}

 \gdef\fullpaper{\gdef\@fullpaper{}}

 \def\affil#1{\\{\small#1\par}}
 \gdef\@author{John Doe1\affil{No-Name University, Shipping Dept.}}
 \long\def\author#1{\gdef\@author{#1}}

 \gdef\@abstract{}
 \long\def\abstract#1{\gdef\@abstract{#1}}

\def\@maketitle{\newpage\leavevmode
  \begin{minipage}[t]{0.30\textwidth}
    \hrule height0pt
    \raggedright
    \mbox{}\par
    \@proofbox
  \end{minipage}\relax
  \begin{minipage}[t]{0.70\textwidth}
    \hrule height0pt
    \raggedleft
    \LARGE\@title\par
    \vskip4pt
    \large\@author
  \end{minipage}
  \vskip8pt
  \ifx\@abstract\@empty\else{\vskip.5em\leftskip1.5in\parskip4pt\small\@abstract\par\vskip.5em}\fi
  \rule{\textwidth}{0.4pt}
  \vskip16pt}

\DeclareRobustCommand\em
        {\@nomath\em \ifdim \fontdimen\@ne\font >\z@
                       \upshape \else \slshape \fi}

\def\@begintheorem#1#2{\sl \trivlist \item[\hskip \labelsep{\bf #1\ #2}]}
\def\@opargbegintheorem#1#2#3{\sl \trivlist
     \item[\hskip \labelsep{\bf #1\ #2\ (#3)}]}

 \newcommand{\sect}[1]{\S\ref{sect.#1}}      

 \newcommand{\sectlabel}[1]{\label{sect.#1}}

  \def\@arabic#1{\number #1} 

\long\def\@makecaption#1#2{
	\vskip\abovecaptionskip
	\sbox\@tempboxa{{\small #1: #2}}%
	\ifdim\wd\@tempboxa>\hsize
	    {\small #1: #2\par}
	\else
	   \global\@minipagefalse
	   \hbox to\hsize{\hfil\box\@tempboxa\hfil}
	\fi
	\vskip \belowcaptionskip}

\def\figstrut#1{\hbox to\linewidth{\vrule height#1\hfill}}

\def\comppad{\thinspace}
\def\comp{\comppad\begingroup \tt \let\do\@makeother \dospecials 
          \@ifstar{\@scomp}{\@comp}}
\def\@scomp#1{\def\@tempa ##1#1{##1\endgroup\comppad}\@tempa}
\def\@comp{\obeyspaces \frenchspacing \@scomp}

\makeatother

\title{How much does it cost to teleport?}
\author{H. F. Chau and Hoi-Kwong Lo \thanks{Supported by DOE grant
 DE-FG02-90ER40542} \affil{
 School of Natural Sciences, Institute for Advanced Study, \\ Olden Lane,
 Princeton, NJ 08540}
}
\date{10 May 1996\\ IASSNS-HEP-96/51}
\abstract{
We show that the entropy of entanglement of a state characterizes its ability
to teleport. In particular, in order to teleport faithfully an unknown quantum
$N$-state, the two users must share an entangled state with at least $\log_2 N$
bits entropy of entanglement. We also note that the maximum capacity for a
mixed state ${\cal M}$ to teleport equals the maximum amount of entanglement
entropy that can be distilled out from ${\cal M}$. Our result, therefore,
provides an alternative interpretation for entanglement purification.
}
\begin{document}
\maketitle
\section{Introduction}\sectlabel{S:Intro}
 In the process called quantum teleportation, an unknown quantum state is
 disassembled into, and later reconstructed from purely classical information
 and purely non-classical EPR correlations. Bennett {\em et al.} showed that
 two bits of classical information and one maximally entangled EPR
 pair are {\it sufficient} for the faithful teleportation of an unknown
 two-state quantum system \cite{Tele93}. They also showed how a simple
 modification of their method can be used to teleport an $N$-state object with
 the resources of $2\log_2 N$ bits of classical information and a pair of
 $N$-state particles in a completely entangled state shared by the two users.
 We shall briefly describe the teleportation scheme introduced by Bennett
 {\em et al.} in \sect{S:Tele}.
\par
 There are a number of conceptually important questions on quantum
 teleportation that we would like to answer. First, what are the minimal
 resources needed
 for a general teleportation scheme? A partial answer was given in
 Ref.~\cite{Tele93} where Bennett {\em et al.} proved that the impossibility
 of superluminal
 communication implies that the reliable teleportation of an $N$-state would
 require a {\it classical} channel of $2 \log_2 N$ bits. In spite of the recent
 advances in quantum information theory (see for example
 \cite{CHB:96,Fuchs,Llyod,Sch:96}), the issue of the {\it minimal} amount of
 {\it quantum} sources required to teleport an $N$-state object has never been
 answered directly.\footnote{One may answer this question
 by using the theorem on
 p.14 of Ref~\cite{CHB:96}, and we shall return to this point later in
 \sect{S:Ent}.}
 Using the idea of Hilbert space dimension counting, we give a
 simple proof
 that, for the faithful
 teleportation of an $N$-state object, the two users must share no less than
 $\log_2 N$ bits entropy of entanglement in \sect{S:Ent}. Consequently, the
 entropy of
 entanglement of a quantum state can also be interpreted as a measure of the
 usefulness of that state in teleportation.
\par
 The second question that we would like to address concerns the capability of
 faithful teleportation using mixed states.
 Owing to the interactions with
 the surroundings, the entangled quantum state shared by Alice and Bob
 before the teleportation should, in general, be a mixed
 state. Faithful teleportation is nonetheless possible:
 Using the so-called entanglement purification
 scheme \cite{Increase_Ent,Noisy_Tele,CHB:96}, one can distill out some
 maximally entangled states. These distilled states can then be used in
 faithful teleportation. In \sect{S:Pure}, we note that the maximum capability
 for a mixed state ${\cal M}$ to teleport is equal to the maximum amount of
 entanglement entropy that can be distilled out from ${\cal M}$.
 Finally, a summary in given in \sect{S:End}.
\section{Teleportation scheme of Bennett {\em et al.}} \sectlabel{S:Tele}
 Teleportation is a method of indirectly sending a quantum state from one place
 to another. Conventionally, the sender is called Alice and the receiver is
 called Bob. As we shall discuss below, Alice sends Bob two messages: a quantum
 message at any time before the actual teleportation, and a classical message
 during the actual teleportation. Teleportation, as opposed to directly sending
 the quantum particle, is preferred when the quantum channel between Alice
 and Bob at the time of the quantum data transfer is jammed or noisy.
\par
 Let us first consider a simple example of teleporting a two-state particle.
 In order to teleport an unknown state in the form $\alpha |0\rangle + \beta
 |1\rangle$ from Alice to Bob, they perform the following operations
 \cite{Tele93}:
\begin{enumerate}
\item
 Alice prepares an EPR singlet. She sends one of the EPR particle to Bob
 through a quantum channel (and for the time being, we assume the channel is
 noiseless). 
 She retains the second EPR particle for herself.
\item 
 Alice makes sure that Bob has received the EPR particle.
\item
 Alice makes a joint measurement on the combined system of the EPR particle
 that she retains and the unknown
 quantum state that she wants to teleport along the four Bell basis, namely,
 $|\Psi^\pm \rangle = \left( |10\rangle \pm |01\rangle \right) / \sqrt{2}$ and
 $|\Phi^\pm \rangle = \left( |11\rangle \pm |00\rangle \right) / \sqrt{2}$.
\item
 She tells Bob the result of her measurement via a classical communication
 channel (and once again, we assume the classical channel is noiseless).
\item
 Bob reconstructs the original unknown state by applying an unitary
 transformation $U$ to his EPR particle according to the measurement result of
 Alice that he receives from the classical channel. In fact, $U = -I$,
 $-\sigma_3$,
 $\sigma_1$, and $i\sigma_2$ if the measurement result of Alice is $|\Psi^{-}
 \rangle$, $|\Psi^{+}\rangle$, $|\Phi^{-}\rangle$, and $|\Phi^{+}\rangle$
 respectively, where $\sigma_i$'s are the Pauli spin matrices.
\end{enumerate}
\par\medskip\indent
 Because of the linearity of quantum mechanics, teleportation of $n$ two-state
 particles can be achieved by teleporting the particles one by one. Besides,
 it is obvious that the above method can be used to teleport both pure and
 mixed states.
\par
 The above scheme can be generalized for the teleportation of an $N$-state
 particle using $2
 \log_2 N$ classical bits of communication plus $\log_2 N$ EPR pairs.
 But, can we use resources fewer than the above scheme? In Ref.~\cite{Tele93},
 Bennett {\em et al.} argued that teleportation of an $N$ state particle using
 less than $2\log_2 N$ classical bits of communication would violate causality.
 In the coming section, we show that at least $\log_2 N$ bits of entanglement
 entropy is also required. Thus, the Bennett {\em et al.} scheme is optimal.
\section{Minimum quantum resources required in teleportation} \sectlabel{S:Ent}
 We give the definition of entanglement entropy before discussing the amount of
 quantum resources needed for teleportation. Actually, there are a number of
 inequivalent definitions for the entanglement entropy of formation for a
 mixed quantum
 state. And here in definition 3, we use the one proposed by Bennett
 {\em et al.} in
 Ref.~\cite{CHB:96}.
\par\medskip\noindent
{\bf Definition~1:} The entanglement entropy $E (|\Psi\rangle )$ of a pure
 state $|\Psi\rangle$
 shared between two parties, Alice and Bob, is defined as the von Neumann
 entropy $S(\mbox{Tr}_{\rm Alice} |\Psi\rangle \langle\Psi |)$ of the mixed
 state which it appears to Alice (or Bob) to be.
\par\medskip\noindent
{\bf Definition~2:} The entanglement entropy of an ensemble of pure states
 ${\cal E} = \{ p_i, |\Psi_i \rangle \}$ shared between Alice and Bob is
 defined as the ensemble average $\sum_i p_i E(|\Psi_i \rangle )$ of the
 entanglement entropy of the pure states in the ensemble.
\par\medskip\noindent
{\bf Definition~3:} The entanglement entropy of formation of a mixed state
 ${\cal M}$ is the
 minimum of $E({\cal E})$ over ensembles ${\cal E}$ realizing the mixed state
 ${\cal M}$.
\par\medskip
 The resources used by the two users, Alice and Bob, may be decomposed into two
 (namely classical and quantum) parts: In addition to some un-entangled
 states that they may possess individually, they also share an
 entangled pair of $M$-state objects which may be completely or partially
 entangled. The idea of our proof on the minimum resources required in
 teleportation is very simple. First, we show that, for the
 faithful teleportation of an $N$-state object, $M$ must be larger than or
 equal to $N$. Second, assuming that quantum teleportation of an $N$-state
 object can be achieved with less than $\log_2 N$ bits entropy of entanglement,
 we show that the condition $M \geq N$ will be violated. Therefore, to avoid
 contradiction, it must be the case that at least $\log_2 N$ bits entropy of
 entanglement are needed for teleporting an $N$-state object.
 An alternative proof based on the idea \cite{Increase_Ent,CHB:96}
 that local operations never increase
 the entropy of entanglement between Alice and Bob can be made.
 However, we stick to the following proof since it is conceptually
 simpler and interesting in its own right. 
\par\medskip\noindent
{\bf Lemma~1:} In order to reliably teleport an $N$-state quantum object, Alice
 and Bob must share an entangled pair of $M$-state objects with $M \geq N$.
\par\noindent
{\it Proof:} We prove by contradiction. Let us assume that Alice and Bob can
 succeed in reliably teleporting the state $| \phi \rangle$ of an unknown
 $N$-state object with an entangled pair of $M$-state objects where $M <N$.
 During the teleportation process, Bob must be able to reconstruct the state
 $| \phi \rangle$. Let $| \Psi_{\rm Bob} \rangle$ denote the quantum state of
 the $M$-state particle in his share after Alice's measurement. Of course, the
 state $| \Psi_{\rm Bob} \rangle$ depends on the outcome of Alice's measurement
 which is also communicated to Bob. In addition, Bob may process some auxiliary
 particles in a state denoted by $|\Psi_{\rm Aux} \rangle$ which is independent
 of the result of Alice's measurement and the state $|\phi\rangle$ of the
 object to be teleported. Bob must then apply an unitary transformation
 $U^{\rm result}$ to $|\Psi_{\rm Bob} \rangle \otimes | \Psi_{\rm Aux} \rangle$
 to reconstruct the state $|\phi\rangle$. The unitary operator
 $U^{\rm result}$ is a function of the result of Alice's measurement. But now
 for {\it any} Alice's measurement result, the support of Bob's constructed
 state, $U^{\rm result} | \Psi_{\rm Bob} \rangle \otimes | \Psi_{\rm Aux}
 \rangle$, is at most $M$-dimensional. Since by assumption $M < N$, Bob will
 clearly fail in reconstructing the original state $|\phi\rangle$: If Bob were
 to succeed in such a reconstruction, transmission of any $N$-state quantum
 object could be decomposed into the transmission of a $M$-state quantum object
 ($M < N$) plus some classical bits! This is clearly impossible.
\hfill $\Box$
\par\medskip\noindent
{\bf Theorem~1:} In order to teleport an unknown $N$-state quantum object,
 Alice and Bob must share an entangled quantum state with entropy of
 entanglement $E \geq \log_2 N$ bits.
\par\noindent
{\it Proof:} Let us assume that quantum teleportation of an unknown $N$-state
 object can be achieved with $E ~(<\log_2 N) $ bits entropy of entanglement
 shared between Alice and Bob. By separately applying quantum data compression
 to their respective subsystems, Alice and Bob could squeeze the original
 entanglement into a smaller number of shared pairs of qubits
 \cite{Jozsa,Sch:95}. Using this
 two-sided compression for $r$ shared pairs of entanglement $E$, Alice and Bob
 will each possess slightly more than $rE$ qubits having slightly less than
 $rE$ bits entropy of entanglement. The state of those slightly larger than
 $rE$ qubits will be an excellent approximation of that of the original $rM$
 qubits and can be used for an almost faithful teleportation of an $N^r$-state
 object. But those slightly larger than $rE$ qubits have a total Hilbert space
 dimension less than $N^r$ and yet they are supposed to be sufficient for the
 reliable teleportation of an $N^r$-state object. This contradicts Lemma~1.
\hfill $\Box$
\par\medskip
 Consequently, if Alice uses $n$ pairs of entangled qubits to teleport an
 unknown state of $n$ qubits, the $n$ pairs of qubits she has to prepare
 must be maximally entangled. In
 this respect, the teleportation scheme proposed by Bennett {\em et al.}
 through an ideal channel in
 Ref.~\cite{Tele93} is optimal since it requires the least possible amount of
 entanglement entropy shared between Alice and Bob.
\par
 The following theorem tells us that the bound given in Theorem~1 is tight:
\par\medskip\noindent
{\bf Theorem~2:} Given a pure state $|\Psi\rangle$ shared between two
 parties. It can be used to teleport $E(|\Psi\rangle )$ qubits.
\par\noindent
{\it Proof:} Given a sufficiently large
 number of copies of $|\Psi\rangle$, Alice and Bob can apply an entanglement
 purification scheme (details of the schemes can be found, for example, in
 Refs.~\cite{Increase_Ent,Noisy_Tele,CHB:96}). Since $|\Psi\rangle$ is a pure
 state, the maximum number of EPR pairs that can be distilled out per copy of
 $|\Psi\rangle$ equals $E(|\Psi\rangle )$ \cite{CHB:96,Sch:96}.
 Then using the Bennett {\em et al.}'s teleportation scheme, we succeed in
 teleporting $E(|\Psi\rangle )$ qubits using $|\Psi\rangle$.
\hfill $\Box$
\par\medskip
 Thus, for a given pure state shared between two parties, the entanglement
 entropy of that state can be interpreted as a measure of the maximum
 capability of the state as an agent for teleportation.
 Nevertheless, the following example
 shows that knowing only the entanglement entropy of the quantum state
 share between Alice and Bob alone is not sufficient to carry out faithful
 teleportation.
\par\medskip\noindent
{\bf Example~1:} Consider the teleportation scheme proposed by Bennett {\em et
 al.} in Ref.~\cite{Tele93}. But instead of using an EPR singlet, Alice and Bob
 share the state $|\Psi\rangle = (|11\rangle + |00\rangle ) / \sqrt{2}$.
 Clearly the entropy of entanglement for $|\Psi\rangle$ equals one bit.
 Therefore, Theorem~1 tells us that it can be used to teleport one qubit from
 Alice to Bob. However, following the procedure of Bennett {\em et al.}, if
 Alice prepares a state $|\psi\rangle = \alpha |0\rangle + \beta |1\rangle$,
 then after the teleportation, Bob gets $|\psi'\rangle =
 -\beta |0\rangle + \alpha |1\rangle$, which is orthogonal to $|\psi\rangle$.
 The reason why this teleportation scheme fails completely is that Bob applies
 an incorrect unitary operation on his quantum particle after getting the
 classical message from Alice. In fact, the unitary operation that Bob needs to
 perform on his quantum particles depends on which entangled state Alice and
 Bob share, not just the amount of entropy entanglement between them.
\par\medskip
 Further discussions on the conditions on the joint measurement needed for
 faithful teleportation can be found in Ref.~\cite{Unpub}.
\section{Equivalence of the purification and teleportation capability}
\sectlabel{S:Pure}
 So far, our discussion is restricted to the case where the quantum
 state shared by Alice and Bob is pure. But in real life,  decoherence occurs
 when the quantum particles are transmitted through a noisy channel. Thus, the
 quantum particles shared by Alice and Bob should be described by a mixed
 state. Given a mixed state ${\cal M}$, the two users are still able to
 perform faithful teleportation. Suppose Alice and Bob share a number
 of identical copies of the mixed states. By means of measurements on some of
 their quantum particles together with some classical communications between
 them, they can distill out a smaller set of quantum particles
 which are maximally entangled. Such schemes are called entangled
 purification protocols, and we refer to
 Refs.~\cite{Increase_Ent,Noisy_Tele,CHB:96} for their detailed procedures.
 Now Alice and Bob carry out faithful teleportation using the purified
 maximally entangled states.
 In what follows, we prove that for a given mixed state ${\cal M}$, the above
 scheme for
 faithful teleportation is already the most efficient one.
 Now, we state some useful definition before giving our proof.
\par\medskip\noindent
{\bf Definition~4:} Let ${\cal M}$ be a mixed state shared between Alice and
 Bob. $D_{{\rm A}\rightarrow {\rm B}} ({\cal M})$ denotes the maximum amount
 of entanglement entropy that can be distilled from ${\cal M}$ by entanglement
 purification protocols which allow only one-way classical communications from
 Alice to
 Bob. $D_{{\rm B} \rightarrow {\rm A}} ({\cal M})$ is defined in a similar way.
 In addition, $D_{{\rm A}\leftrightarrow {\rm B}} ({\cal M})$ denotes the
 maximum amount of entanglement entropy that can be distilled from ${\cal M}$
 by entanglement purification protocols which allow two-way classical
 communications between both Alice and Bob.
\par\medskip\noindent
{\bf Definition~5:} Let ${\cal M}$ be a mixed state shared between Alice and
 Bob. $T_{{\rm A}\rightarrow {\rm B}} ({\cal M})$ denotes the maximum amount
 of qubits that Alice can faithfully teleport, given that only one-way
 classical
 communications from Alice to Bob is allowed. Similarly, $T_{{\rm A}
 \leftrightarrow {\rm B}} ({\cal M})$ denotes the maximum amount of qubits
 that Alice
 can faithfully teleport when two-way classical communications between both
 Alice and Bob is allowed.
\par\medskip\noindent
{\bf Remark~1:} It is easy to see that $E ({\cal M}) \geq D_{{\rm A}
 \leftrightarrow {\rm B}} ({\cal M}) \geq D_{{\rm A}\rightarrow {\rm B}}
 ({\cal M})$ and $E ({\cal M}) \geq T_{{\rm A}\leftrightarrow {\rm B}}
 ({\cal M}) \geq T_{{\rm A}\rightarrow {\rm B}} ({\cal M})$ for any mixed state
 ${\cal M}$. And as shown in Ref.~\cite{CHB:96}, there are situations where
 $D_{{\rm A}\leftrightarrow {\rm B}}$ is strictly greater than $D_{{\rm A}
 \rightarrow {\rm B}}$.
\par\medskip\noindent
{\bf Theorem~3:} $T_{{\rm A}\rightarrow {\rm B}} = D_{{\rm A}\rightarrow
 {\rm B}}$ and $T_{{\rm A}\leftrightarrow {\rm B}} = T_{{\rm A}\leftrightarrow
 {\rm B}}$.
\par\noindent
{\it Proof:} We only prove the first equality. The proof of the second equality
 is similar to the first one. First, we show that $T_{{\rm A}\rightarrow
 {\rm B}} \geq D_{{\rm A}\rightarrow {\rm B}}$: For a given mixed state
 ${\cal M}$, we use the optimal purification scheme, which allows only
 classical communications from Alice to Bob, to give $D_{{\rm A}\rightarrow
 {\rm B}} ({\cal M})$ maximally entangled pairs per impure pair. Since the
 quantum state of these purified pairs are known, and they are maximally
 entangled, from Theorem~2 and Example~1, we can faithfully teleport
 $D_{{\rm A}\rightarrow {\rm B}} ({\cal M})$ qubits from Alice to Bob using
 only one way classical communication from Alice to Bob. Thus, $T_{{\rm A}
 \rightarrow {\rm B}} \geq D_{{\rm A}\rightarrow {\rm B}}$.
\par
 It remains to show that $D_{{\rm A}\rightarrow {\rm B}} \geq T_{{\rm A}
 \rightarrow {\rm B}}$. Again, we consider a given mixed state ${\cal M}$. By
 means of the optimal teleportation scheme involving only one way classical
 communication from Alice to Bob, we can faithfully teleport $T_{{\rm A}
 \rightarrow {\rm B}} ({\cal M})$ qubits per impure pair. Clearly, Alice can
 prepare some perfectly entangled EPR pairs, and then use the above
 teleportation scheme to faithfully ``transport'' half of her pairs to Bob.
 After this, Alice and Bob are able to share $T_{{\rm A}\rightarrow {\rm B}}
 ({\cal M})$ perfectly entangled EPR pairs per impure pair. Thus, $D_{{\rm A}
 \rightarrow {\rm B}} \geq T_{{\rm A}\rightarrow {\rm B}}$.
\hfill $\Box$
\par\medskip
 Theorem~3 provides an alternative interpretation for $T_{{\rm A}\rightarrow
 {\rm B}}$ and $T_{{\rm A}\leftrightarrow {\rm B}}$. They measure the
 capability of both faithful teleportation and entanglement purification using
 one-way and two-way classical communications, respectively.
\par\medskip
 Recently, using some ideas from teleportation, Bennett {\em et al.} argued
 that entanglement purification schemes
 are closely related to quantum error-correcting code (see for example
 Refs.~\cite{QEC7,QEC4,QEC6,QEC3,QEC5,QEC1,QEC2} for the various quantum
 error-correcting codes proposed). And it is interesting to further investigate
 the relationship between teleportation and quantum error-correcting codes.
\section{Summary} \sectlabel{S:End}
 In summary, we study the cost of a general teleportation scheme. Using a
 simple idea of Hilbert space dimension counting, we prove in \sect{S:Ent} that
 in order to teleport an unknown $N$-state quantum signal, a quantum state with
 entanglement entropy $E$ of at least
 $\log_2 N$ is required to be shared between Alice and Bob.
 Consequently, we conclude that the Bennett {\em et
 al.}'s teleportation scheme via an ideal quantum channel is optimal because it
 uses the minimum possible amount of classical and quantum resources. We also
 argue that the entanglement entropy for a pure state can be interpreted as the
 usefulness of a state in teleportation.
\par
 We go on to consider the case of mixed state. we find that
 the maximum capability of a mixed state ${\cal M}$ to perform
 faithful teleportation is equal to the maximum amount of
 entanglement entropy that can be distilled out from ${\cal M}$.
 This provides, once again, an alternative interpretation of
 $D_{{\rm A}\rightarrow {\rm B}}$ and $D_{{\rm A}\leftrightarrow {\rm B}}$.
\par\medskip
 A number of open questions remain. First, what is the relation between
 entanglement entropy of formation and the maximum amount of qubits that a
 mixed state can faithfully teleport?
 It is quite conceivable that $E ({\cal M}) > T_{{\rm A}\leftrightarrow
 {\rm B}} (
 {\cal M})$ for some mixed state ${\cal M}$, although a rigorous proof is
 lacking (compare with Remark~1). 
 Second, can we characterize the fidelity of teleportation
 when the entangled state shared between the two parties actually differs
 slightly from the one they have in mind?
\par\medskip\noindent
{\it Acknowledgments:}
 One of us (H.F.C.) would like to thank Alexander Korotkov for his discussion
 which eventually evolved to the present work.


\begin{thebibliography}{99}
\bibitem{Tele93} {\sc Bennett}, Charles H., G. {\sc Brassard}, C.
 {\sc Cr\'{e}peau}, R. {\sc Jozsa}, A. {\sc Peres}, and W. K. {\sc Wootters},
 ``Teleporting an unknown quantum state via dual classical and
 Einstein-Podolsky-Rosen channels'', {\em Phys. Rev. Lett.} {\bf 70} (1993),
 1895-1899.
\bibitem{Increase_Ent} {\sc Bennett}, Charles H., H. J. {\sc Bernstein}, S.
 {\sc Popescu}, and B. {\sc Schumacher}, ``Concentrating partial entanglement
 by local operations'', {\em Phys. Rev. A \bf 53}, (1996) 2046-2052.
\bibitem{Noisy_Tele} {\sc Bennett}, Charles H., G. {\sc Brassard}, S.
 {\sc Popescu}, B. {\sc Schumacher}, J. A. {\sc Smolin}, and W. K.
 {\sc Wootters}, ``Purification of noisy entanglement and faithful
 teleportation via noisy channels'', {\em Phys. Rev. Lett.} {\bf 76} (1996),
 722-725.
\bibitem{CHB:96} {\sc Bennett}, Charles H., D. P. {\sc DiVincenzo}, J. A.
 {\sc Smolin}, and W. K. {\sc Wootters}, ``Mixed state entanglement and quantum
 error correction'', Los Alamos preprint archive {\tt quant-ph/9604024} (Apr,
 1996).
\bibitem{QEC7} {\sc Calderbank}, A. Robert, and P. W. {\sc Shor}, ``Good
 quantum error-correcting codes exist'', Los Alamos preprint archive
 {\tt quant-ph/9512032} (Dec, 1995).
\bibitem{Unpub} {\sc Chau}, Hoi Fung, and H.-K. {\sc Lo}, ``Quantum
 teleportation using EPR pairs is optimal and unique'', unpublished technical
 report (Mar, 1996).
\bibitem{QEC4} {\sc Ekert}, Artur and C. {\sc Macchiavello}, ``Error correction
 in quantum communication'', Los Alamos preprint archive {\tt quant-ph/9602022}
 (Feb, 1996).
\bibitem{Fuchs} {\sc Fuchs}, Christopher A., {\em Distinguishability and
 accessible information in quantum theory}, Ph.D. thesis, Univ. of New Mexico
 (Dec, 1995).
\bibitem{Jozsa} {\sc Jozsa}, Richard, and B. {\sc Schumacher}, ``A new proof of
 the quantum noiseless coding theorem'', {\em J. Mod. Optics} {\bf 41} (1994),
 2343-2349.
\bibitem{QEC6} {\sc Knill}, Emanuel, and R. {\sc Laflamme}, ``A theory of
 quantum error-correcting codes'', Los Alamos preprint archive
 {\tt quant-ph/9604034} (Apr, 1996).
\bibitem{QEC3} {\sc Laflamme}, Raymond, C. {\sc Miquel}, J. P. {\sc Paz}, and
 W. H. {\sc Zurek}, ``Perfect quantum error correction code'', Los Alamos
 preprint archive {\tt quant-ph/9602019} (Feb, 1996).
\bibitem{Llyod} {\sc Llyod}, Seth, ``The capacity of the noisy quantum
 channel'', Los Alamos preprint archive {\tt quant-ph/9604015} (Apr, 1996).
\bibitem{QEC5} {\sc Plenio}, M. B., V. {\sc Vedral}, and P. L. {\sc Knight},
 ``Optimal realistic quantum error correction code'', Los Alamos preprint
 archive {\tt quant-ph/9603022} (Mar, 1996).
\bibitem{Sch:95} {\sc Schumacher}, Benjamin, ``Quantum coding'', {\em Phys.
 Rev. A} {\bf 51} (1995), 2738-2747.
\bibitem{Sch:96} {\sc Schumacher}, Benjamin, ``Sending entanglement through
 noisy quantum channels'', Los Alamos preprint archive {\tt quant-ph/9604023}
 (Apr, 1996).
\bibitem{QEC1} {\sc Shor}, Peter W., ``Scheme for reducing decoherence in
 quantum memory'', {\em Phys. Rev. A} {\bf 52} (1995), 2493-2496.
\bibitem{QEC2} {\sc Steane}, Andrew, ``Multiple particle interference and
 quantum error correction'', Los Alamos preprint archive {\tt quant-ph/9601029}
 (Oct, 1995).
\end{thebibliography}
\end{document}